\documentclass[rmp,aps,preprint,nofootinbib,endfloats]{revtex4}
\usepackage{graphicx}
\usepackage{epsfig}
\usepackage{amsmath}
\usepackage{bm}
\def\inbar{\,\vrule height1.5ex width.4pt depth0pt}
\def\IR{\relax{\rm I\kern-.18em R}}
\def\IC{\relax\hbox{$\inbar\kern-.3em{\rm C}$}}

\begin{document}


\title{Creating traveling waves from standing waves from the gyrotropic 
paramagnetic properties of Fe$^{3+}$ ions in a high-Q whispering gallery 
mode sapphire resonator}

\author{Karim Benmessai}
\email{kbenmess@cyllene.uwa.edu.au}
\affiliation{University of Western Australia, School of Physics M013, 
35 Stirling Hwy., Crawley 6009 WA, Australia}
\author{Michael Edmund Tobar}
\email{mike@physics.uwa.edu.au}
\affiliation{University of Western Australia, School of Physics M013, 
35 Stirling Hwy., Crawley 6009 WA, Australia}
\author{Nicholas Bazin}
\affiliation{Institut FEMTO-ST, UMR 6174 CNRS, Universit\'e de Franche-Comt\'e, 
25044 Besan\c con, France}
\author{Pierre-Yves Bourgeois}
\affiliation{Institut FEMTO-ST, UMR 6174 CNRS, Universit\'e de Franche-Comt\'e, 
25044 Besan\c con, France}
\author{Yann Kersal\'e}
\affiliation{Institut FEMTO-ST, UMR 6174 CNRS, Universit\'e de Franche-Comt\'e, 
25044 Besan\c con, France}
\author{Vincent Giordano}
\affiliation{Institut FEMTO-ST, UMR 6174 CNRS, Universit\'e de Franche-Comt\'e, 
25044 Besan\c con, France}

\date{\today}

\begin{abstract}  
We report observations of the gyrotropic change in magnetic susceptibility 
of the Fe$^{3+}$ electron paramagnetic resonance at 12.037GHz (between spin 
states $|1/2>$ and $|3/2>$) in sapphire with respect to applied magnetic field. 
Measurements were made by observing the response of the high-Q Whispering 
Gallery doublet (WGH$_{\pm17,0,0}$) in a Hemex sapphire resonator cooled to 5 K. 
The doublets initially existed as standing waves at zero field and were 
transformed to traveling waves due to the gyrotropic response.
\end{abstract}                                                                 


\maketitle


\section{INTRODUCTION}
\label{sec:intro}

Residual paramagnetic impurities in ultra-pure Hemex sapphire crystals greatly 
influence the electromagnetic properties of the resonators cooled to near liquid
helium temperature. For example, the electron paramagnetic resonance (EPR) of 
impurities such as Cr$^{3+}$, Ti$^{3+}$ and Mo$^{3+}$, influence the temperature 
dependence and allow frequency-temperature turnover (annulment) near liquid 
helium temperature, which can vary between 2 to 10 
K\cite{JonesBlair1988el, Mann1992jpDap, Luiten1996jpDap, Kovacich1997jpDap}. 
This phenomenon is vital to produce ultra-stable frequencies for cryogenic 
sapphire oscillators\cite{Giles1990PhysicaB, Pyb2004el, Hartenett2006apl, Locke2006rsi} 
and the study of EPR in low-loss crystals has become an important topic of 
investigation\cite{Hartnet1999uffc, Hartnet2000fmc, Hartnet2007prB, Hartnet2001jpDap}. 
More recently, it was shown that the small amount of residual Fe$^{3+}$ ions 
(less than parts per million) are sufficient to create Maser oscillations due 
to the zero-field 3-level system between $|1/2>$, $|3/2>$ and $|5/2>$ spin 
states\cite{Pyb2005apl, Pyb2006ijmpB, Benmessai2007el}. This has allowed a new 
way to operate a cryogenic sapphire oscillator, where population inversion is 
created by pumping the spin states in the $|1/2>$ transition to the $|5/2>$ 
transitions with a 31.3 GHz pump, and has the potential to operate with a 
frequency stability governed by the Schawlow-Townes noise 
limit\cite{Benmessai2008prl} (sub $10^{-16}$ frequency stability).

In this work we show how the application of an axial magnetic field on a 
cryogenic sapphire resonator, with a mode tuned on the Fe$^{3+}$ EPR, adds a 
gyrotropic component of magnetic susceptibility. This is achieved due to the 
very high-Q whispering gallery (WG) modes (of order $10^9$) that are excited, 
which enables the discrimination of the two degenerate standing wave modes known 
as a doublet. Precision measurement of the frequency and Q-factor of the doublet 
reveals an asymmetric response, which demonstrates that the susceptibility added 
by the magnetic field is predominately of gyrotropic origin and that the modes 
transform to traveling waves. Traveling waves are known to enhance the 
interaction between the pump radiation and the maser medium due to the 
elimination of standing wave nodes, and should be a way of generating greater 
output power in the respective solid-state maser.

\section{WHISPERING GALLERY MODE FIELDS}
\label{sec:WGMF}

\begin{figure}[ht!!] 
\epsfysize=3in 
\epsfbox{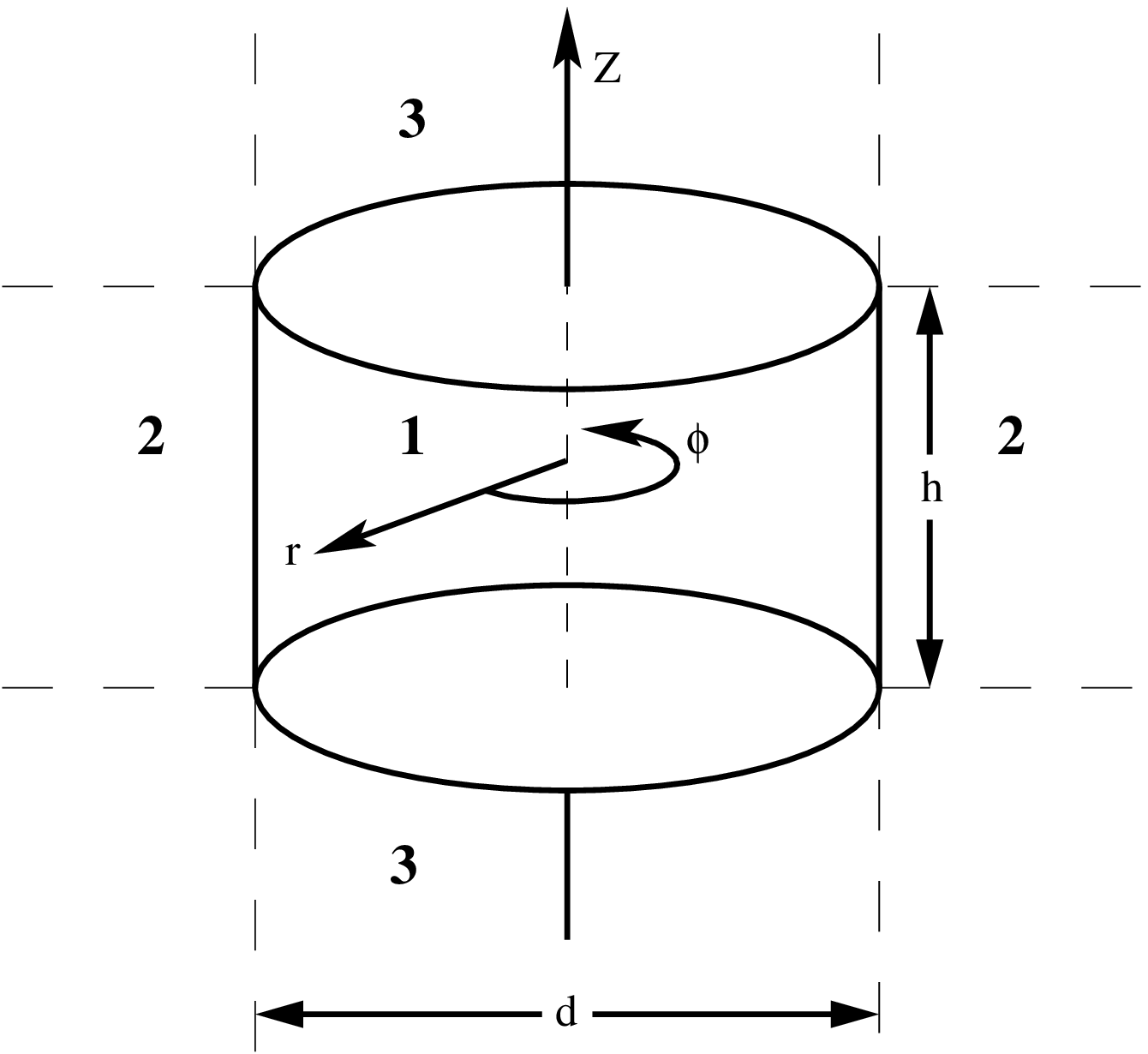} 
\caption{ Solution regions for separation of variables in cylindrical 
coordinates. \label{Fig:Figure1}} 
\end{figure}

To determine the electromagnetic fields of the mode under investigation we use a 
separation of variables technique, which has proved to be reliable for the 
approximate calculation of electromagnetic fields and properties of high-Q WG 
modes in low loss sapphire resonators\cite{TobarMann1991mtt, Wolf2004grg}. To 
calculate the gyrotropic properties we apply perturbation analysis on the 
non-gyrotropic fields. The technique assumes the resonator is a perfect cylinder 
of uniaxial anisotropy with the c-axis aligned along the cylinder z-axis as 
shown in Fig. \ref{Fig:Figure1}.

For electromagnetic modes exisiting in non-gyrotropic media with one or more 
azimuthal variations ($m>0$), two degenerate mode solutions exist 
(doublet) as orthogonal displaced standing wave solutions proportional to either 
$cos(m\phi)$ or $sin(m\phi)$. The degeneracy of the modes is lifted by a very 
small amount (only observable in high-Q systems) due to dielectric backscatter 
or other perturbations, which couple the counter clockwise (CCW, $e^{jm\phi}$) 
or clockwise (CW, $e^{-jm\phi}$) parts between the two doublet standing wave 
modes\cite{Weiss1995ol, Kippenberg2002ol, Gorodetsky2000josaBop}. In gyrotropic 
media, the doublets have their degeneracy further modified in a non-reciprocal 
fashion, and the modes must transform to the travelling wave basis (rather than 
standing wave) with opposite elliptical polarization due to the non-reciprocity 
imposed by the gyrotropic effect\cite{Krupka1996mtt, Krupka1996mttbis}.

In this work we observe gyrotropic effects on a quasi-Transverse Magnetic (TM) 
WG doublet pair, which is represented by the notation, WGH$_{\pm m,n,p}$. Here 
$n$ the number of radial zero crossings and $p$ the number of axial zero 
crossings and the '$\pm$' symbol represents the sense of the elliptical 
polarization of the doublets. For the fundamental WGH$_{\pm m,0,0}$ mode family 
the E$_\textrm{z}$ electric field is the dominant component and is symmetric 
along the z-axis (around z = 0). Given these conditions one can show that the 
solutions to Maxwell's equations yield (the time dependence $e^{j\omega t}$ is 
assumed):
	\begin{equation}\label{eqn:MaxwellSolution1}
	\begin{split}
	E_{z1\pm} & =  A_{1\pm} e^{\pm jm \phi} cos(\beta z) J_m (k_E r)
	 \\ 
	E_{z2\pm} & =  A_{2\pm} e^{\pm jm \phi} cos(\beta z) K_m (k_{out} r)
	 \\ 
	E_{z3\pm} & =  A_{3\pm} e^{\pm jm \phi} e^{-\alpha_E z} J_m (k_E r)
	 \\ 
	H_{z1\pm} & =  B_{1\pm} e^{\pm jm \phi} sin(\beta z) J_m (k_H r)
	 \\
	H_{z2\pm} & =  B_{2\pm} e^{\pm jm \phi} sin(\beta z) K_m (k_{out} r)
	 \\
	H_{z3\pm} & =  B_{2\pm} e^{\pm jm \phi} e^{-\alpha_H z} J_m (k_{H} r)
	\end{split}
	\end{equation}
To calculate the other components we apply the following Maxwell's relationships 
to the z-components given in Eq. (\ref{eqn:MaxwellSolution1}) in the regions 1, 2 
and 3 (derived from Maxwell's equations in cylindrical co-ordinates).
\begin{subequations} \label{eqn:2}
\begin{equation} \label{eqn:2a}
\left( k^2+ \dfrac{\partial ^2}{\partial z^2} \right) H_{r\pm}=j\omega 
\varepsilon_\bot \varepsilon_0 \dfrac{1}{r}\dfrac{\partial E_{z\pm}}{\partial 
\phi}+ \dfrac{\partial^2 H_{z\pm}}{\partial z\partial r}
\end{equation}
\begin{equation} \label{eqn:2b}
\left( k^2+ \dfrac{\partial ^2}{\partial z^2} \right) E_{\phi\pm}=j\omega \mu_0 
\dfrac{\partial H_{z\pm}}{\partial r}+ \dfrac{1}{r} \dfrac{\partial^2 E_{z\pm}} 
{\partial z\partial \phi}
\end{equation}
\begin{equation} \label{eqn:2c}
\left( k^2+ \dfrac{\partial ^2}{\partial z^2} \right) H_{\phi\pm}=-j\omega 
\varepsilon_\bot 
\varepsilon_0 \dfrac{1}{r}\dfrac{\partial E_{z\pm}}{\partial \phi}+ 
\dfrac{\partial^2 H_{z\pm}}{\partial z\partial \phi}
\end{equation}
\begin{equation} \label{eqn:2d}
\left( k^2+ \dfrac{\partial ^2}{\partial z^2} \right) E_{r\pm}=-j\omega \mu_0 
\dfrac{\partial H_{z\pm}}{\partial r}+ \dfrac{1}{r}\dfrac{\partial^2 E_{z\pm}} 
{\partial z\partial r}
\end{equation}
\end{subequations}
Here, $\varepsilon_{//}=$ 11.349 and $\varepsilon_\bot=$ 9.272, are the uniaxial 
permittivity components of sapphire cooled close to liquid helium temperature 
($\sim$5 K) parallel and perpendicular to the c-axis respectively.

The mode pair that we are interested in is the WGH$_{\pm 17,0,0}$ inside a 
sapphire crystal of 50 mm diameter and 30 mm height at room temperature. 
Contraction from room temperature to 5 K is taken into account in the modelling 
by the factors of 0.99927 along the z-axis and 0.99939 perpendicular to the 
z-axis\cite{Tobar1997jpDap, Krupka1999mst, Krupka1999mtt}. Matching boundary 
conditions with the same approximations as given in\cite{Wolf2004grg}, we obtain 
the following values of the parameters as shown in Table \ref{tab:Table1}; 
note that initially no gyrotropic effects are assumed (the effects are small 
and perturbation analysis suffices). This means the initial calculation of 
the + and - polarized modes are degenerate.

\begin{table}
\caption{Calculated parameters from matching boundary conditions for the 
WGH$_{\pm 17,0,0}$ mode pair.}
\label{tab:Table1}
\begin{tabular}{rrrrrrr}
Frequency$\pm$	&	$k_H$	&	$k_E$	&	$\beta$	&	
$k_{out}$	&	$\alpha_H$	&	$\alpha_E$\\ \hline

12.0305 GHz	&	717.707	&	802.952	&	103.866	&	
760.709		&	841.609		&	j 229.754\\ 
\end{tabular}
\end{table}

The parameters in Table \ref{tab:Table1} are independent of the elliptical 
polarization, and electromagnetic filling factors may be calculated from 
Eq.(\ref{eqn:pmipme}), which also turn out to be independent of elliptical 
polarization, and are shown in Table \ref{tab:Table2}.
\begin{equation}\label{eqn:pmipme}
p_{mi}=\dfrac{\mathop{\iiint}_{V_i}\mu_0 H_i \cdot H_i d\nu}{\mathop{\iiint 
}_{V}\mu_0 H \cdot H d\nu} p_{ei} = \dfrac{\mathop{\iiint}_{V_i}\varepsilon_0 
E_i \cdot E_i d\nu}{\mathop{\iiint}_{V} \varepsilon(\nu) E \cdot E d\nu}
\end{equation}
Here $i$ defines the component of the field and corresponding component of 
filling factor within the corresponding region of volume $V_i$. The 
normalization factor calculates the total field energy across all regions in the 
total volume of the resonator, $V$.
\begin{table}
\caption{Electromagnetic filling factors within sapphire dielectric (region 1) 
for the WGH$_{\pm 17,0,0}$ mode pair.}
\label{tab:Table2}
\begin{tabular}{rrrrrr}
$p_{ez1}$	&	$p_{er1}$	&	$p_{e\phi1}$	&	
$p_{mz1}$	&	$p_{mr1}$	&	$p_{m\phi1}$\\\hline
0.970906	&	0.00342936	&	0.0188647	&	
0.0029268	&	0.835142	&	0.0790977\\ 
\end{tabular}
\end{table}

Note that the largest components of the field within the sapphire are $E_z$ and 
$H_r$, this means that the dominant propagation is around the azimuth (Poynting 
Theorem) as expected for a WG mode. Also, the three most dominant components 
($H_r$, $H_\phi$ and $E_z$), are that of a TM mode with the $H_\phi$ component 
combining with the $H_r$ to define the sense of the elliptical polarization. The 
amount of field in the vacuum surrounding the dielectric (regions 2 and 3) may 
be calculate by, $1-p_{ez}-p_{er}-p_{e\phi}=$ 0.00679954, and $1-p_{mz}-p_{mr}- 
p_{m\phi}=$ 0.0828332 respectively.

To determine the sense of elliptical polarization we take a closer look at the 
solution of the dominant components for the + and - mode in region 1 (the 
sapphire dielectric), which are calculated by substituting Eq. (\ref{eqn:MaxwellSolution1}) 
into (\ref{eqn:2}):
	\begin{equation}\label{eqn:MaxwellSolution2}
	\begin{split}
	E_{z1+} & = e^{jm \phi} cos(\beta z) \left[ -E_{z1o} J_m (k_E r) \right]
	 \\ 
	H_{r1+} & = e^{jm \phi} cos(\beta z) \left[ H_{r1a}\dfrac{\partial 
	J_m(k_H r)}{\partial r} + H_{r1b} \dfrac{J_m(k_E r)}{r}\right]
	 \\ 
	H_{\phi1+} & =j e^{jm \phi} cos(\beta z) \left[ H_{\phi 1a}\dfrac{\partial 
	J_m(k_H r)}{\partial r} + H_{\phi 1b} \dfrac{J_m(k_E r)}{r}\right]
	 \\ 
	E_{z1-} & = e^{-jm \phi} cos(\beta z)  \left[ E_{z1o} J_m (k_H r) \right]
	 \\
	H_{r1-} & = e^{-jm \phi} cos(\beta z) \left[ H_{r1a}\dfrac{\partial 
	J_m(k_H r)}{\partial r} + H_{r1b} \dfrac{J_m(k_E r)}{r}\right]
	 \\
	H_{\phi1-} & = -je^{-jm \phi} cos(\beta z) \left[ H_{\phi 
	1a}\dfrac{\partial J_m(k_H r)}{\partial r} + H_{\phi 1b} 
	\dfrac{J_m(k_E r)}{r}\right]
	\end{split}
	\end{equation}
Here, the solution requires $H_{r1a} = 1.09\times10^{-7}E_{z1o}$, $H_{r1b} = 
1.82\times10^{-4}E_{z1o}$, $H_{\phi1a} = 1.86\times10^{-6}E_{z1o}$, $H_{\phi1b} 
= 1.07\times10^{-5}E_{z1o}$. Thus, by inspection the polarization of the 
magnetic field for the CCW propagating mode is proportional to $e^{jm\phi}$ and 
has CCW elliptical polarization (+ mode), while the magnetic field for the CW 
propagating mode proportional to $e^{-jm\phi}$ has CW elliptical polarization 
(- mode).

\section{GYROTROPIC REPRESENTATION}
\label{sec:Gyro}

In this work we show that gyrotropic effects occur on the WGH$_{17,0,0}$ 
doublet pair when tuned within the Fe$^{3+}$ electron paramagnetic resonance in 
the presence of a DC magnetic field. To calculate the gyrotropic nature we use 
the same notation as presented by Gurevich\cite{Gurevich1960sppl}. We assume the 
case of a weak magnetic response, which is true for our situation as we use the 
purest HEMEX sapphire available, which has no more than the order of parts per 
million of residual impurity ions \cite{JonesBlair1988el, Mann1992jpDap, Hartnet2007prB, Tobar2003prD}. 
In this case the magnetic susceptibility added can be considered independent of 
the microwave field, i.e. only dependent on the applied DC field. The 
permeability added by the electron paramagnetic resonance response to the 
magnetic field in the axial direction, is a tensor of the form:
\begin{equation}\label{eqn:5}
\overleftrightarrow{\mu}(B)=\left[ 
\begin{array}{ccc}
\mu_\perp (B)	&	j\mu_a(B)	&	0\\
-j\mu_a(B)	&	\mu_\perp (B)	&	0\\
0	&	0	&	1
\end{array} 
\right]
\end{equation}
Thus, the change in permeability (or change in susceptibility) created by the 
response to the external field is given by the following:
\begin{equation}\label{eqn:6}
\Delta \overleftrightarrow{\mu}(B)= \overleftrightarrow{\mu}(B) -\mu_0 
\overleftrightarrow{I} =
\left[ 
\begin{array}{ccc}
\Delta \mu_\perp (B)	&	j\mu_a(B)	&	0\\
-j\mu_a(B)	&	\Delta \mu_\perp (B)	&	0\\
0	&	0	&	0
\end{array} 
\right]
\end{equation}
Here, $\overleftrightarrow{I}$ is the unitary matrix and in general the 
permeability components are complex.

\section{APPLICATION OF PERTURBATION THEORY}
\label{sec:Pert}

The typical value of susceptibility added by residual impurities is of the order 
$10^{-8}$, thus we apply perturbation theory based on the fields calculated 
using separation of variables in Sec \ref{sec:Gyro}. Following the method in 
Gurevich, the change in complex frequency maybe calculated from the change in 
complex permeability from,
\begin{equation}\label{eqn:7}
\dfrac{\Delta f_{\pm}}{f_{\pm}}=\dfrac{\mathop{\iiint}_{V_i}H_{\pm}^*\Delta 
\overleftrightarrow{\mu}H_{\pm}d\nu}{\iiint_V \varepsilon(\nu)E_{\pm} 
\cdot E_{\pm}d\nu +\mathop{\iiint}_{V} \mu_0 H_{\pm}\cdot H_{\pm} d\nu}
\end{equation}
Here, $f_0$ is the unperturbed frequency (no gyrotropic effect) so that $\Delta 
f_{_\pm} = f_{_\pm} - f_0$, where $\Delta f_{_\pm}/f_{_\pm}$ is in general complex. In this 
work the complex fractional frequency is defined by $\Delta f_{_\pm} /f_{_\pm} =\Delta 
f_{Re_{\pm}} /f_{_\pm} + j \Delta f_{Im_\pm} /f_{_\pm}$ , where the real part describes 
the frequency shift and the imaginary part the unloaded Q shift, where Q$^{-1}_ 
{\textrm{B}_\pm} \approx 2\Delta f_{Im_\pm}/f_{_\pm}$. The total mode unloaded Q 
values (Q$_{0_\pm}$) is related to the zero field unloaded Q value (Q$_{ZF_\pm}$) 
and the shift due to the magnetic field by Q$^{-1}_{0_\pm}$=Q$^{-1}_{ZF_\pm}+$Q$^ 
{-1}_{\textrm{B}_\pm}$. For the above model the complex permeability added by the 
magnetic field is of the form $\Delta \mu_\perp = \Delta \mu_\perp '-j 
\Delta \mu_\perp ''$ and $\mu_a = \mu_a '-j\mu_a ''$. 
Substituting the complex permeability into Eq. (\ref{eqn:6}), and then (\ref{eqn:6}) 
into (\ref{eqn:7}) along with the unperturbed fields, the following 
sensitivities with respect to the permeability are calculated for the WGH$_{ 
\pm 17,0,0}$ doublet.
	\begin{equation}\label{eqn:Deltanu/nu}
	\begin{split}
	\dfrac{\Delta f_{Re+}}{f_+} &=-0.457176(\Delta \mu_\perp 
	'-0.10485 \mu_a ') = -0.457176 \Delta 
	\mu_{eff+} '
	 \\ 
	\dfrac{\Delta f_{Re-}}{f_-} &=-0.457176(\Delta \mu_\perp 
	'+0.10485 \mu_a ') = -0.457176 \Delta 
	\mu_{eff-} '	 
	 \\ 
	\dfrac{\Delta f_{Im+}}{f_+} &=0.457176(\Delta \mu_\perp 
	''-0.10485 \mu_a '') = 0.457176 \Delta 
	\mu_{eff+} ''=\dfrac{1}{2\textrm{Q}_{\textrm{B}+}}	 
	\\
	\dfrac{\Delta f_{Im-}}{f_-} &=0.457176(\Delta \mu_\perp 
	''+0.10485 \mu_a '') = 0.457176 \Delta 
	\mu_{eff-} ''=\dfrac{1}{2\textrm{Q}_{\textrm{B}-}}
	\end{split}
	\end{equation}
The non-reciprocity of the two modes is highlighted by Eq. (\ref{eqn:Deltanu/nu}), 
which shows the - and + mode dependent on different effective permeability 
$\Delta \mu_{eff_\pm}= \Delta \mu_{eff_\pm}' -j \Delta \mu_{eff_\pm} ''$ (both 
real and imaginary).

The magnetization model of Gurevich predicts $\mu_\perp '$ and $\mu_a ''$ to be 
of the same sign. Thus, given that the both modes tune up in frequency with 
magnetic field the '-' or CW mode must be the mode with the greatest tuning 
coefficient. According to the model the mode should also be accompanied with a 
decreasing Q-factor, while the '+' or CCW mode increases its Q-factor. This 
is indeed what is observed, in the next section two resonators are analysed and 
the values of $\Delta \mu_\perp '$, $\mu_a '$, $\Delta \mu_\perp '$ and $\mu_a'$ 
calculated as a function of magnetic field.

\section{EXPERIMENTAL RESULTS}
\label{sec:Exp}

\begin{figure} 
\epsfysize=6in 
\epsfbox{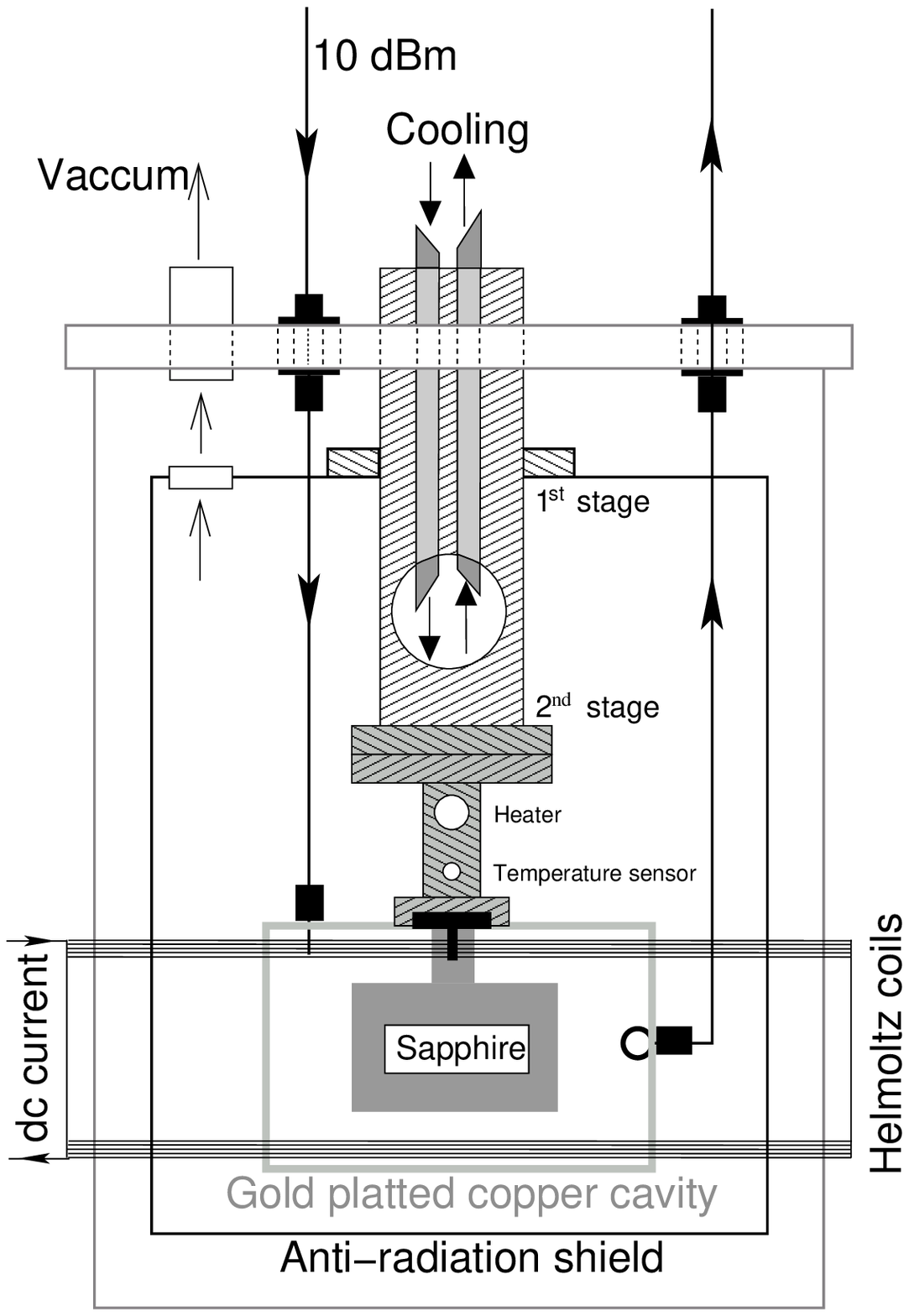} 
\caption{The sapphire resonator was mounted in a gold platted copper cavity and 
measure in transmission with 10 dBm of incident power from a network analyser. 
The DC magnetic field in the axial direction was applied by a Helmholtz coil 
pair system as shown above. The resonator was placed in vacuum and fixed on a 
cryocooler (PT 405) cold finger and cooled close to 5 K using a temperature 
control system. \label{Fig:Figure2}} 
\end{figure}

Hemex sapphire crystals are known to have Fe$^{2+}$ impurities in the crystal. 
To convert these ions to Fe$^{3+}$ the crystals need to be annealed in an oxygen 
environment\cite{Benabid2000}. The resulting ESR exhibits a zero field ESR 
between the $|1/2>$ and $|3/2>$ spin states at 12.037 GHz with a 27 MHz 
bandwidth\cite{Pyb2005apl, Pyb2006ijmpB, Benmessai2008prl}. The crystal 
resonators are nominally 50 mm diameter and 30 mm in height, for such a 
resonator the WGH$_{\pm 17,0,0}$ doublet pair is tuned inside the ESR bandwidth, 
with a predicted frequency from the separation of variables technique of 12.03 
GHz. In this work we report on the properties of two crystals, which were 
annealed. Measurements were taken on a crystal, which was not annealed, however 
no measurable effects with respect to magnetic field were observed. The 
experimental set up of the cryogenically cooled system with a variable external 
magnetic field is shown in Fig. \ref{Fig:Figure2}. 

\begin{figure} 
\epsfysize=2.5in 
\epsfbox{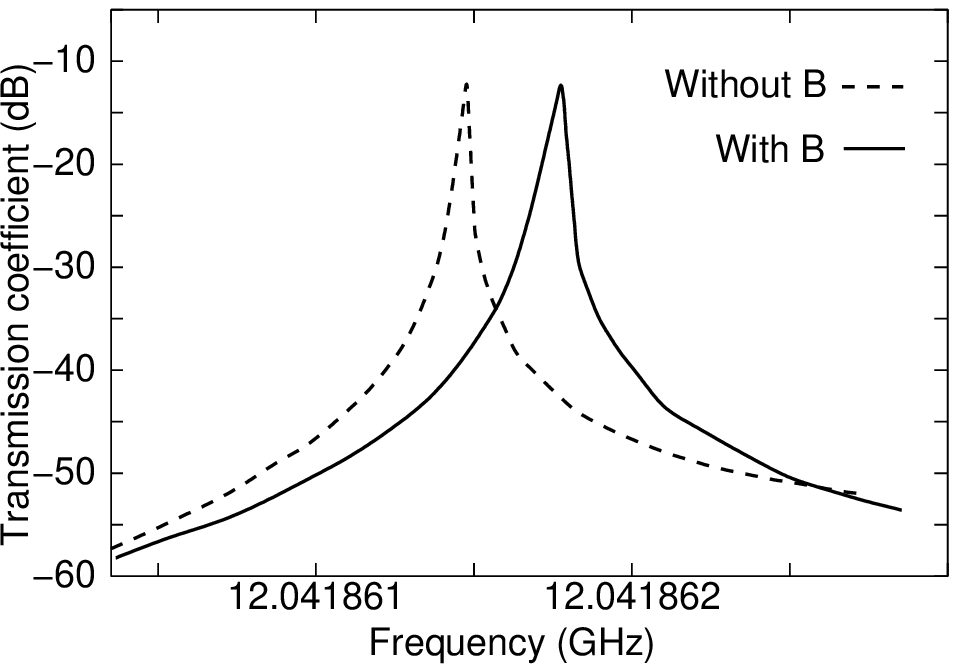} 
\caption{Frequency, Q factor and mode coupling were recorded using a network 
analyser referenced to a Hydrogen maser to avoid frequency drift of the network 
analyser. For resonator 1 the coils sensitivity is 4 gauss/A and 1 gauss/A for 
resonator 2. \label{Fig:Figure3}} 
\end{figure}

The Q-factors and frequencies of the two WGH$_{\pm 17,0,0}$ doublet modes were 
recorded as a function of magnetic field for two separate sapphire resonators 
using a network analyser in transmission. A typical recording of the 
transmission coefficient with and without the magnetic field on is shown in 
Fig. \ref{Fig:Figure3}. 

\begin{figure} 
\epsfysize=2.5in 
\begin{minipage}[c]{0.49\linewidth}
\epsfbox{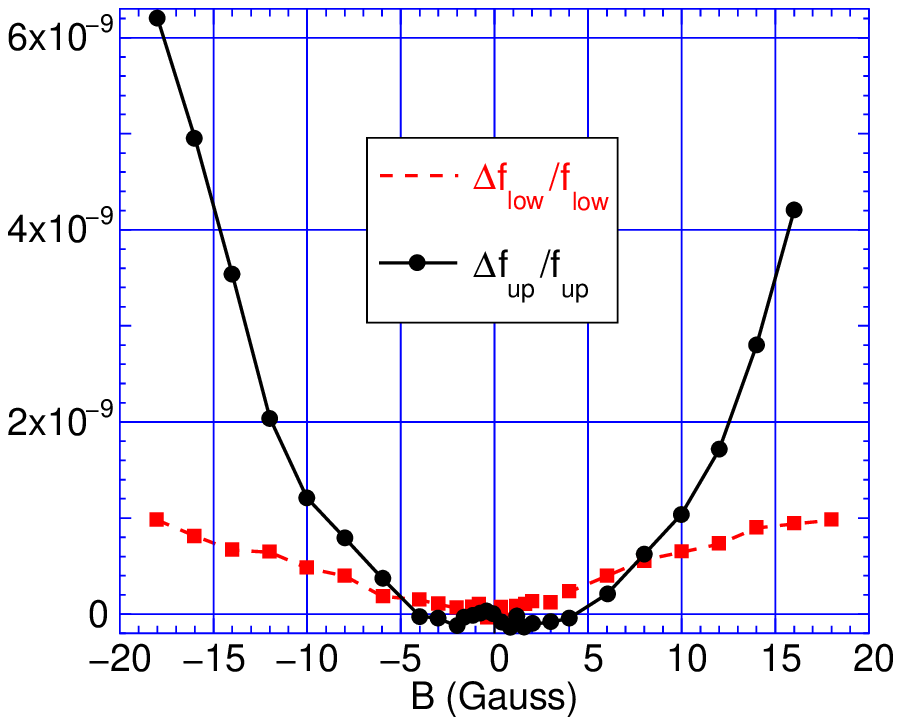} 
\end{minipage}\hfill
\begin{minipage}[c]{0.49\linewidth}
\epsfbox{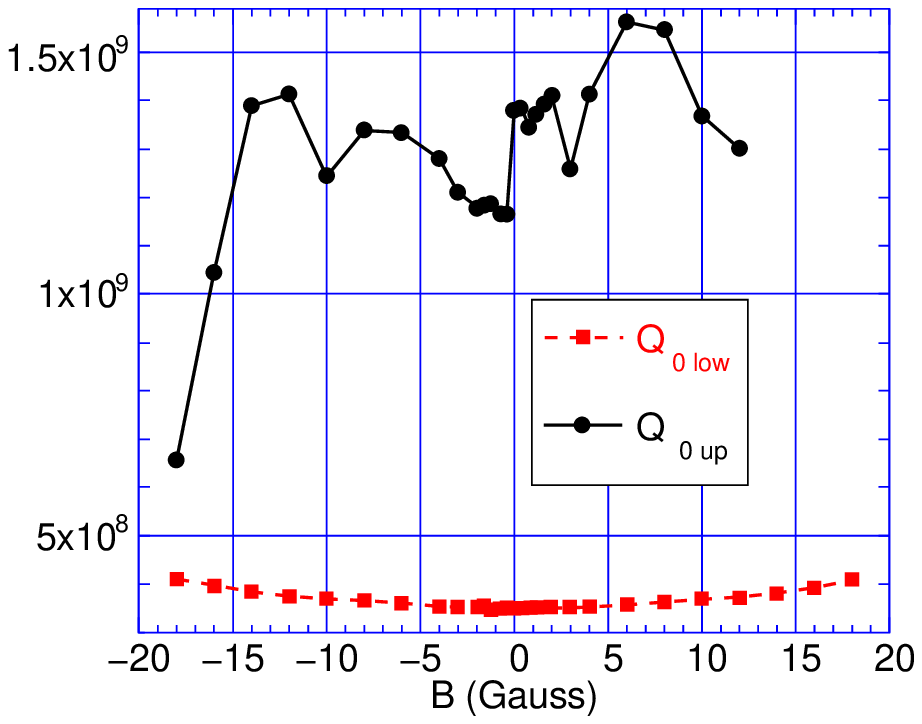} 
\end{minipage}
\caption{Left, Measured fractional frequency shift for the WGH$_{\pm 17,0,0}$ 
lower and upper doublet modes as a function of applied axial magnetic field for 
sapphire No. 1 (Table 3). Right, corresponding measured unloaded Q-factor. 
\label{Fig:Figure4}} 
\end{figure}

\begin{figure} 
\epsfysize=2.5in 
\begin{minipage}[c]{0.49\linewidth}
\epsfbox{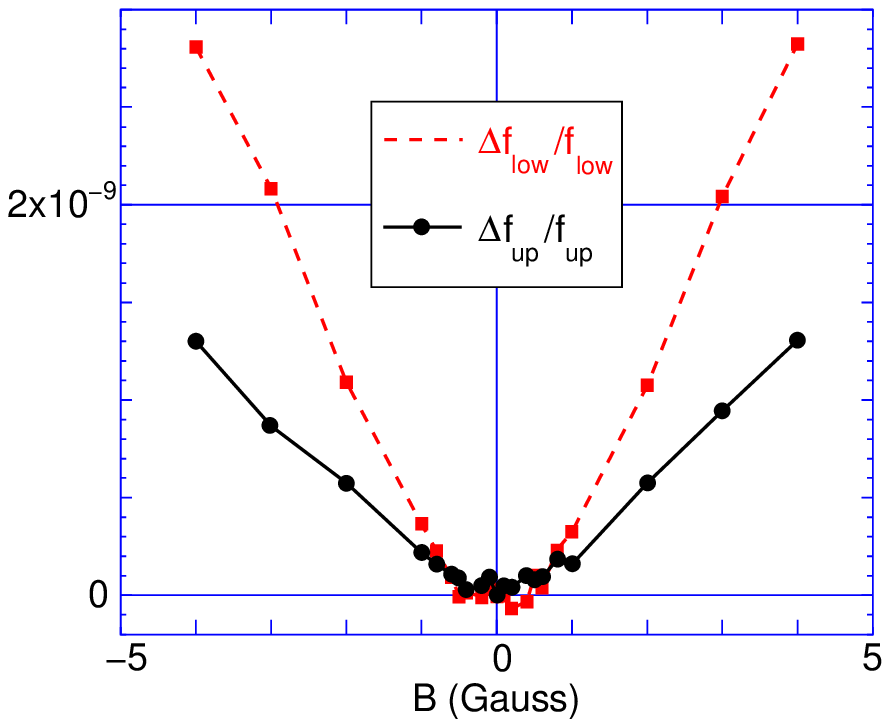} 
\end{minipage}\hfill
\begin{minipage}[c]{0.49\linewidth}
\epsfbox{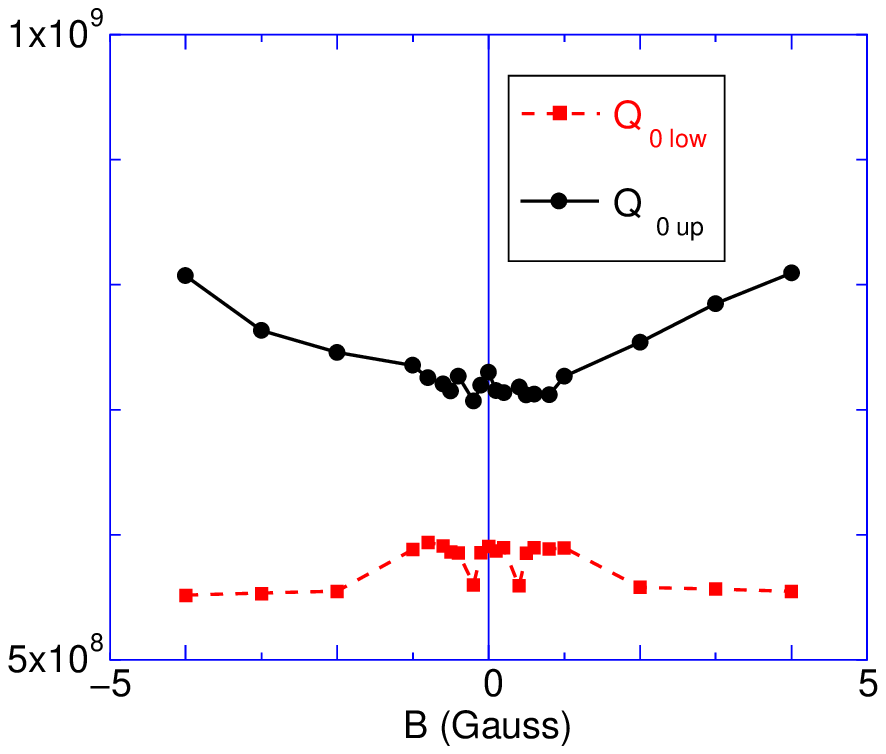} 
\end{minipage}
\caption{Left, Measured fractional frequency shift for the WGH$_{\pm 17,0,0}$ 
lower and upper doublet modes as a function of applied axial magnetic field for 
sapphire No. 2 (Table 4). Right, corresponding measured unloaded Q-factor. 
\label{Fig:Figure5}} 
\end{figure}

With zero field applied to the crystals the following parameters in Tab. 3 and 
4 were measured. The doublet modes were already separated in frequency at 
zero-field due to other non-gyrotropic perturbations\cite{Kippenberg2002ol, Gorodetsky2000josaBop, pyb2005mtt}. 
The modes are labelled as upper (subscript up) or lower (subscript low) for the 
higher and lower frequency tuned modes. The frequency shift and unloaded 
Q-factor of the upper and lower modes were measured as a function of magnetic 
field and are shown in Fig. \ref{Fig:Figure4} and \ref{Fig:Figure5} for the two 
resonators. 

\begin{table}
\caption{Zero field properties of the WGH$_{\pm 17,0,0}$ doublet in sapphire 
resonator number 1.}
\label{tab:Table3}
\begin{tabular}{rrrrrr}
Mode (Sapphire No.1)	&	Frequency [GHz]	&	Q$_0$ \\ \hline
lower	&	12.023984181	&	$3.5\times10^8$\\ 
upper	&	12.023994100	&	$13.8\times10^8$\\ 
\end{tabular}
\end{table}

\begin{table}
\caption{Zero field properties of the WGH$_{\pm 17,0,0}$ doublet in sapphire 
resonator number 2.}
\label{tab:Table4}
\begin{tabular}{rrrrrr}
Mode (Sapphire No.2)	&	Frequency [GHz]	&	Q$_0$ \\ \hline
lower			&	12.041856348	&	$5.9\times10^8$\\ 
upper			&	12.041861773	&	$7.3\times10^8$\\ 
\end{tabular}
\end{table}

\begin{figure} 
\epsfysize=2.5in 
\begin{minipage}[c]{0.49\linewidth}
\epsfbox{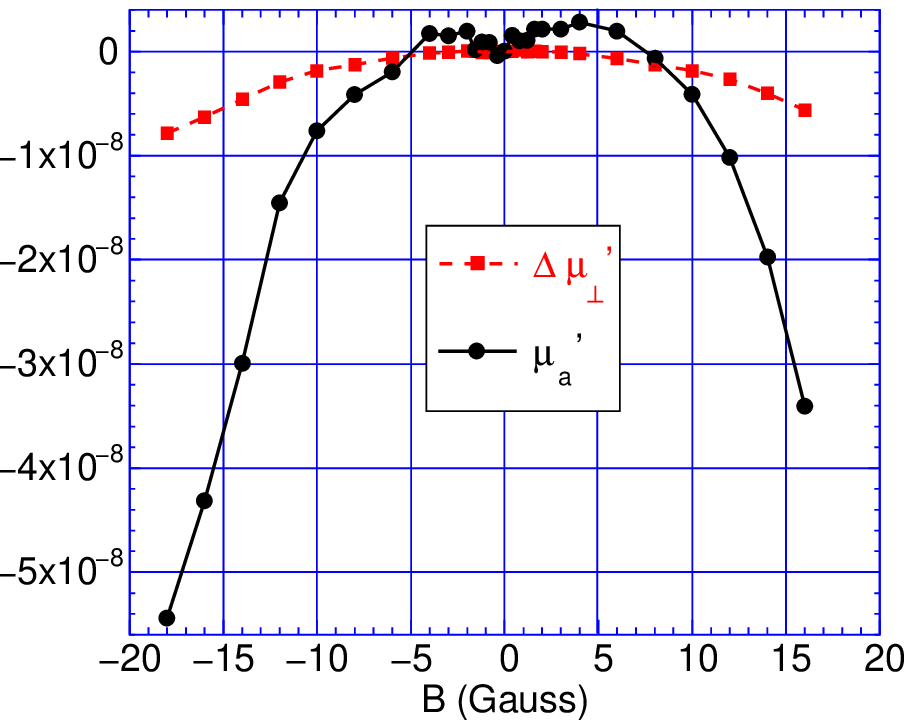} 
\end{minipage}\hfill
\begin{minipage}[c]{0.49\linewidth}
\epsfbox{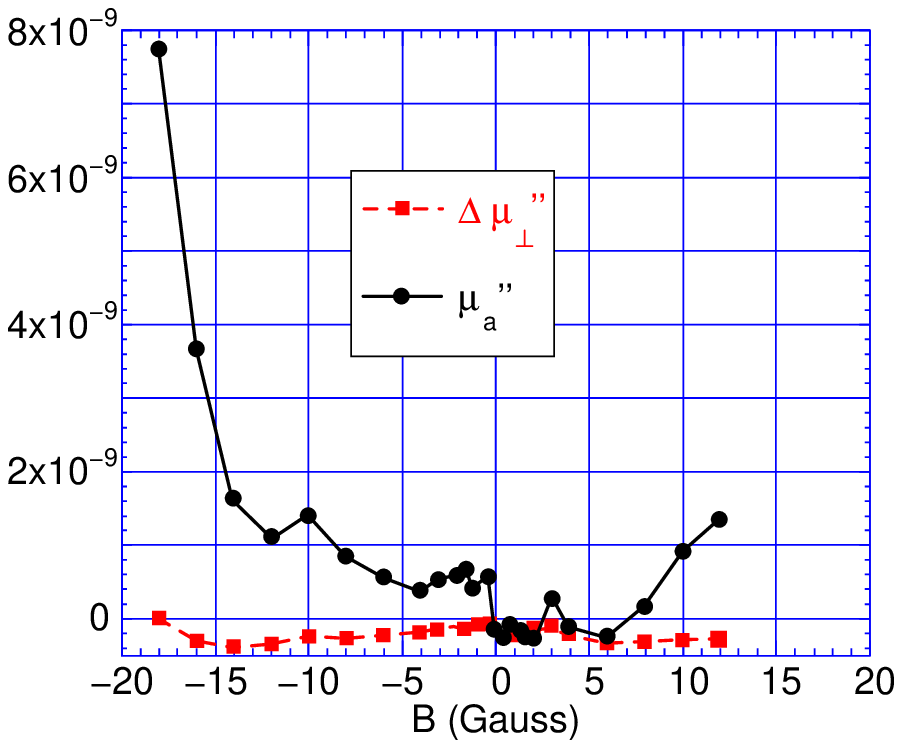} 
\end{minipage}
\caption{Left, Real permeability components as a function of applied axial 
magnetic field for sapphire No. 1. Right, corresponding imaginary components. 
Note, the gyrotropic response is larger in both cases.
\label{Fig:Figure6}} 
\end{figure}

\begin{figure} 
\epsfysize=2.5in 
\begin{minipage}[c]{0.49\linewidth}
\epsfbox{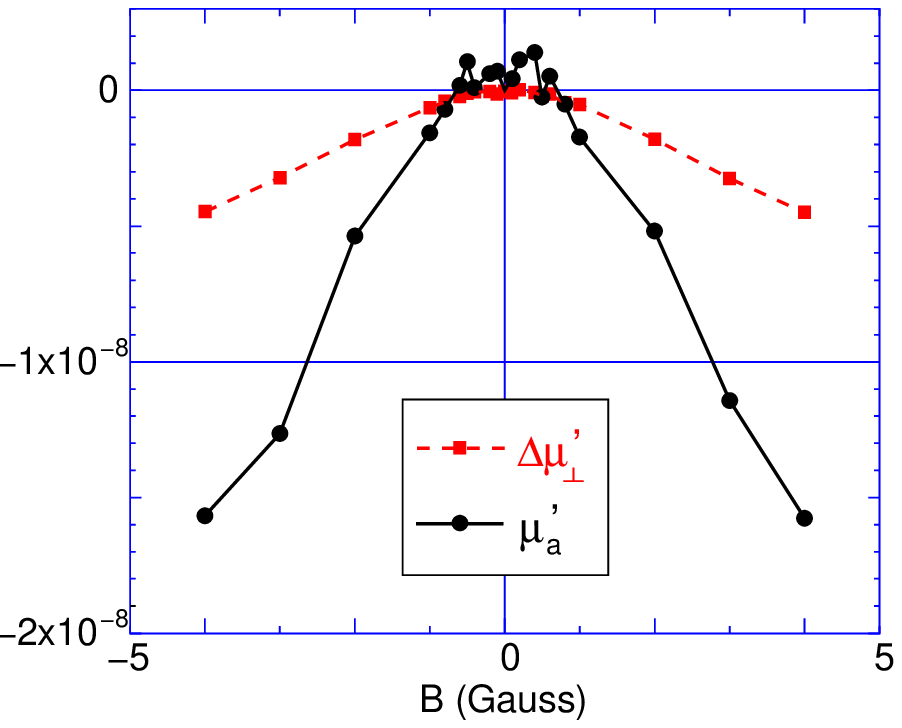} 
\end{minipage}\hfill
\begin{minipage}[c]{0.49\linewidth}
\epsfbox{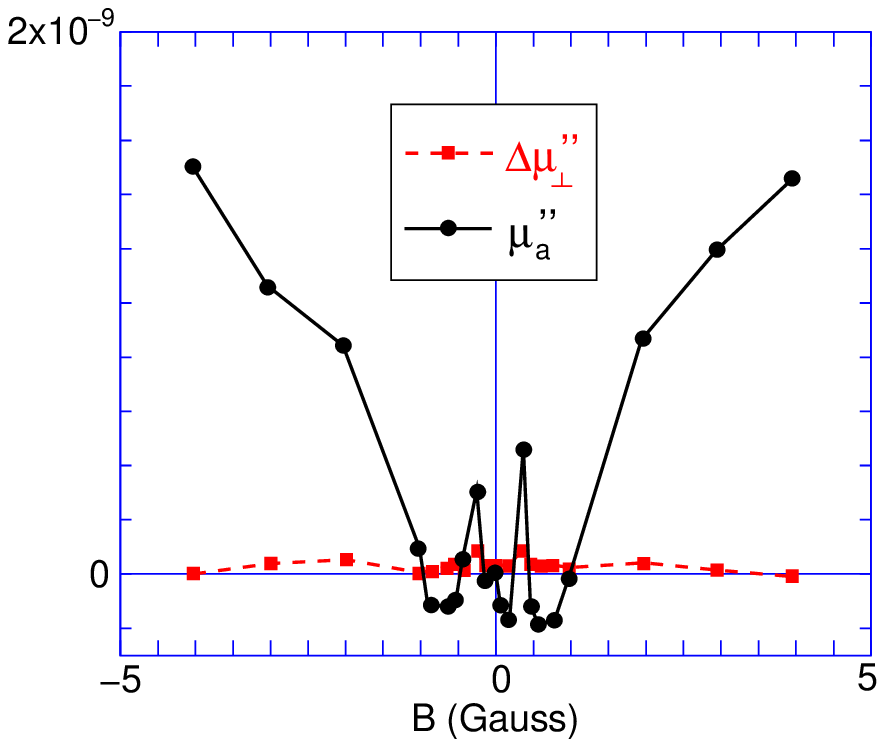} 
\end{minipage} 
\caption{Left, Real permeability components as a function of applied axial 
magnetic field for sapphire No. 2. Right, corresponding imaginary components. 
Note, the gyrotropic response is larger in both cases.
\label{Fig:Figure7}} 
\end{figure}

From the properties present in Fig. \ref{Fig:Figure4} and \ref{Fig:Figure5}, it 
is straightforward to determine the polarizations of the modes. The '-' CW 
polarized mode has a larger effective permeability with respect to magnetic 
field and should thus tune at a greater rate. The same mode should also increase 
its effective loss tangent at the same time. Thus, for sapphire No. 1 the '-' 
mode is identified as the upper mode and the '+' CCW mode as the lower. It is 
the reverse situation for sapphire No. 2 as the lower mode is the '-' mode and 
the upper is the '+'. After this identification, equation \ref{eqn:DeltamuDefinition}) 
may be used to solve for the real and imaginary permeabilities. The solutions 
lead to the following equations:
\begin{equation}\label{eqn:DeltamuDefinition}
	\begin{array}{rrr}
	\Delta \mu_\perp '=-1.09367\left( \dfrac{\Delta f_{\textrm{Re}-}}{f_-} 
	+\dfrac{\Delta f_{\textrm{Re}+}}{f_+} \right) 
	& 
	\Delta \mu_ \perp''=1.09367\left(\dfrac{1}{\textrm{Q}_{\textrm{B}-}}+\dfrac{1} 
	{\textrm{Q}_{\textrm{B}+}}\right)\\
	
	\mu_a '=-10.4305\left( \dfrac{\Delta f_{\textrm{Re}- 
	}}{f_-}-\dfrac{\Delta f_{\textrm{Re}+}}{f_+} \right) 
	& 
	\mu_a 
	''=10.4305\left(\dfrac{1}{\textrm{Q}_{\textrm{B}-}}+\dfrac{1}{\textrm{Q}_{\textrm{B}+}} 
	\right)
	\end{array} 
\end{equation}
Thus, from Eq. (\ref{eqn:DeltamuDefinition}) and the data of Fig. \ref{Fig:Figure4} 
and \ref{Fig:Figure5}, the change in complex permeability with respect to 
magnetic field may be calculated and is plotted in Fig. \ref{Fig:Figure6} and 
\ref{Fig:Figure7}. 

It is clear that the gyrotropic susceptibility added by the magnetic field is 
larger than the non-gyrotropic component for both real and imaginary terms.

\section{TRAVELLING WAVES}
\label{sec:TravWaves}

\begin{figure} 
\epsfysize=2.5in 
\epsfbox{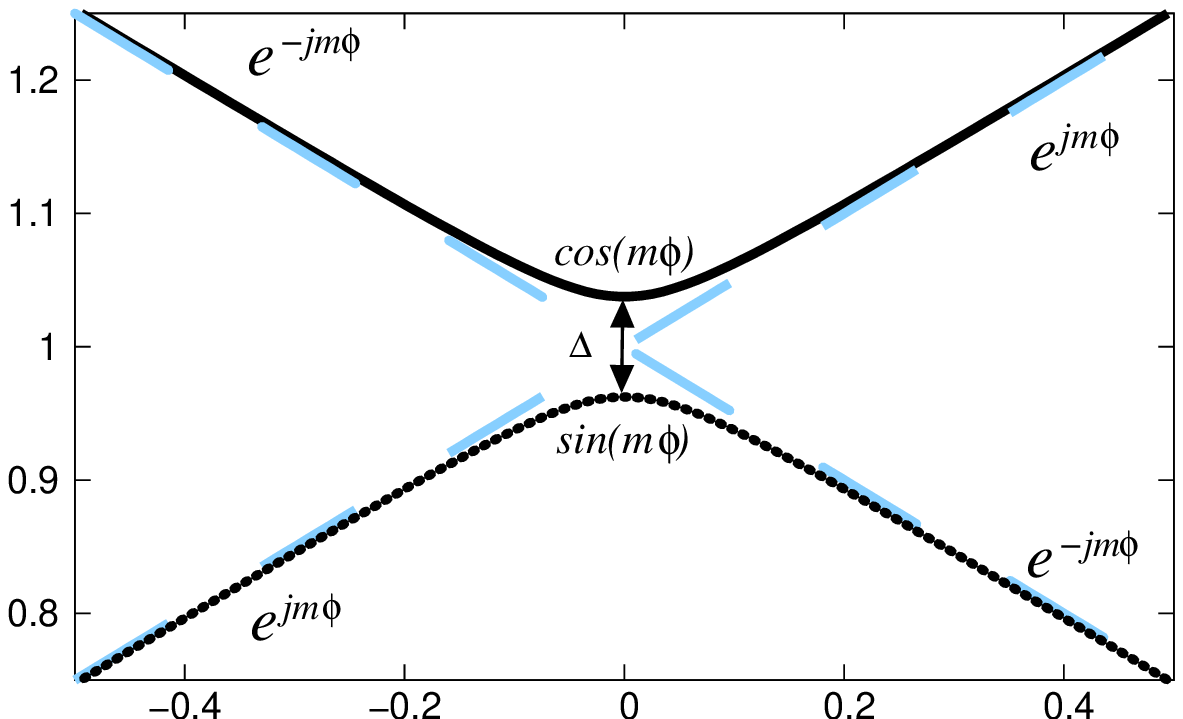} 
\caption{Normal mode frequencies versus tuning parameter, $x$, with $k=1/2$ and 
$\Delta= 0.1$ as given by Eq. (\ref{eqn:fvsx}). The top solid curve is the 
upper tuned mode $f_{up}/f_0$, the bottom solid curve is the lower tuned mode 
$f_{low}/f_0$. The non-coupled modes in the travelling wave basis $f_{_\pm}/f_0$ 
are given by the dashed lines. When the modes are detuned they behave as 
travelling waves, when they are tuned the modes interact and are mixed and are 
best described by standing waves ($cos(m\phi)$ and $sin(m\phi)$). The extent of 
the transition region depends on the strength of the coupling. 
\label{Fig:Figure8}} 
\end{figure}

Coupled mode theory is used to show that standing waves are converted to 
travelling waves. Previous analysis has shown that in high-Q systems 
perturbations, including backscatter due to the dielectric and other 
perturbations such as probes, and will cause the WG doublet to have its degeneracy 
lifted\cite{Weiss1995ol, Gorodetsky2000josaBop, Mazzei2007prl}. Coupled modes 
in tuneable high-Q sapphire resonators have been studied in detail 
previously\cite{Tobar1991mtt, Tobar1993jpDap}. The lifting of the degeneracy is 
caused by reactive coupling through the electromagnetic fields (strong 
coupling), which creates a band gap between the modes. The frequency of the 
upper and lower modes with respect to a linear tuning parameter, $x$, are in 
general given by Eq. (\ref{eqn:fvsx}).
\begin{equation}\label{eqn:fvsx}
	\begin{split}
	\dfrac{f_{up/low}(x)}{f_0}=1\pm\sqrt{k^2x^2+\Delta^2/4}
	\\ 
	\dfrac{f_+(x)}{f_0}=1+kx\; \; \; \; \; \dfrac{f_-(x)}{f_0}=1-kx
	\\ 
	\end{split}
\end{equation}
Here the frequencies of the travelling waves $f_+$ and $f_-$ depend linearly on 
the tuning parameter $x$, with a tuning coefficient of $\pm k$ respectively, an 
idealized diagram with respect to coupled mode theory with $k=1/2$, is shown in 
Fig. \ref{Fig:Figure8}. When the two modes have the same frequency, i.e. $x = 0$ 
the mode separation due to the strong coupling is $f_{up}/f_0 - f_{low}/f_0= 
\Delta$, which defines the strength of the coupling and the band gap width. 

\begin{figure} 
\epsfysize=4in 
\epsfbox{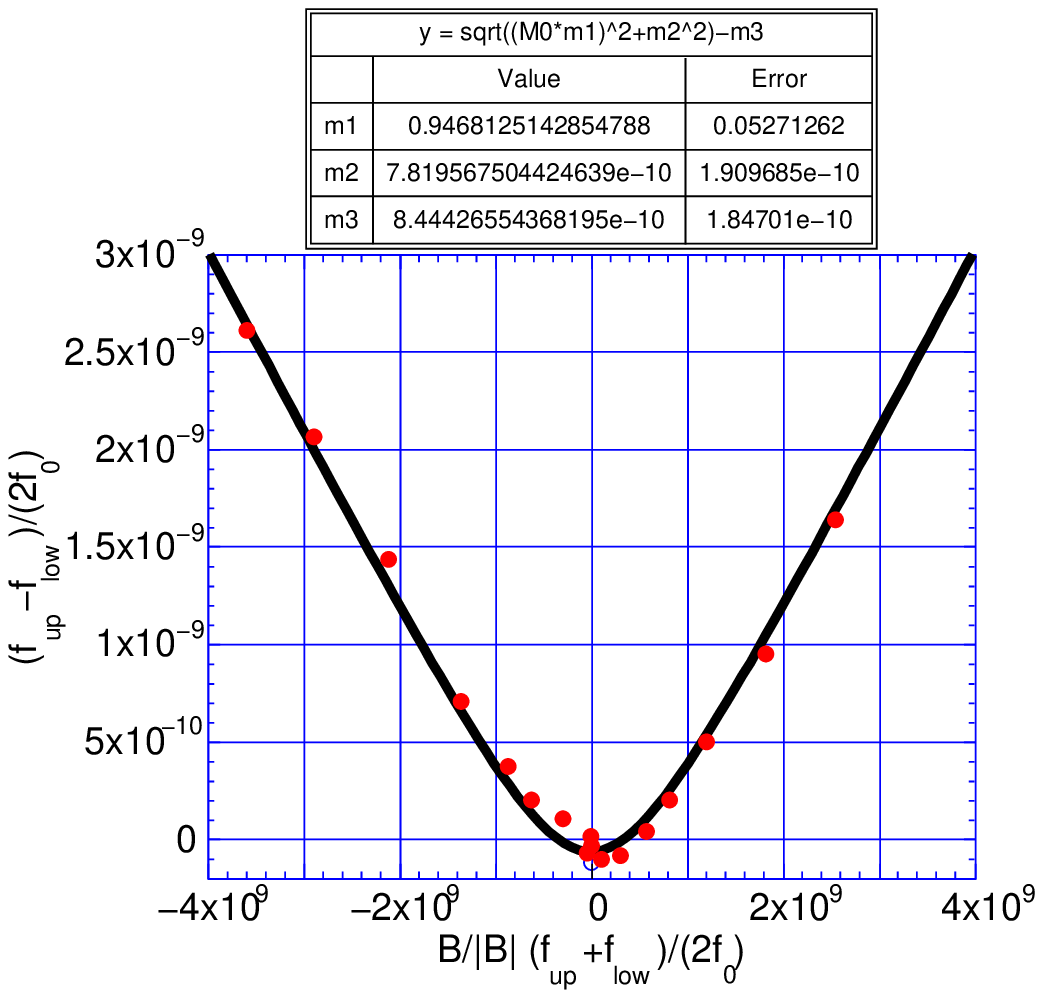} 
\caption{Fractional differential mode frequency shift versus common mode 
frequency shift for sapphire No. 1. The curve fit gives the coupling parameter 
$(m2)$ of $\Delta = 7.8(1.9)\times10^{-10}$ and coefficient $(m1)$ $k=0.95(0.05 
)$.
\label{Fig:Figure9}} 
\end{figure}

\begin{figure} 
\epsfysize=4in 
\epsfbox{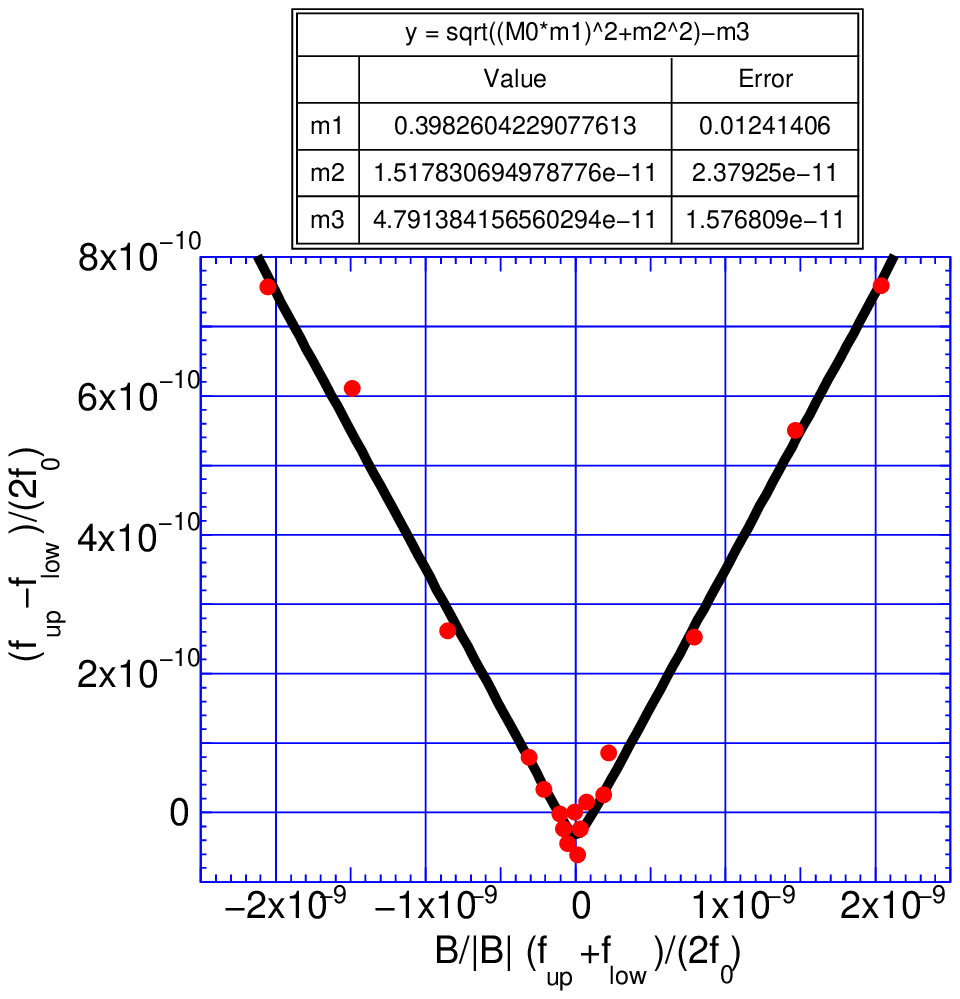} 
\caption{Fractional differential mode frequency shift versus common mode 
frequency shift for sapphire No. 2. The curve fit gives the coupling parameter 
$(m2)$ of $\Delta = 1.5(2.4)\times10^{-11}$ and coefficient $(m1)$ $k=0.40(0.01 
)$.
\label{Fig:Figure10}} 
\end{figure}

The data represented in Fig. \ref{Fig:Figure4} and \ref{Fig:Figure5} shows the 
mode frequencies tuning as a function of magnetic field. The magnetic field 
tunes the frequencies due to the tuning of the EPR, however the tuning is 
non-linear. Given that the non-gyrotropic response of the EPR to the magnetic 
field should be in the linear regime, as the magnetic fields are relatively 
small\cite{Kornienko1961} we may define the average frequency shift of the two 
modes as an effective linear tuning parameter, such that $x=B/|B|(f_{up} + f_{ 
low})/2f_0$. Fig. \ref{Fig:Figure9} and \ref{Fig:Figure10} show the differential 
shift between the two modes as a function of $x$ (ignoring the initial mode 
splitting). The differential shift according to Eq. (\ref{eqn:fvsx}) will be 
of the form $(f_{up} - f_{low})/2f_0 =\sqrt{k^2x^2+\Delta^2/4}$, and may be 
fitted to the experimental data to calculate the coupling parameter, $\Delta$, 
between the travelling waves. 

The ESR is inhomogenously broadened so the interaction will be with the spin 
packet of closest frequency in both sapphires. Sapphire 2 has a lower threshold 
of about 0.6 Gauss compared to sapphire 1 with 4 Gauss. The threshold value 
basically corresponds to the value of magnetic field that the standing waves are 
transformed to travelling waves and the point where the mode responds linearly 
to tuning parameter $x$ with the value of $k$ for the tuning coefficient. The 
threshold magnetic field value also correlates with the value of the coupling 
strength between the travelling waves with the higher value demanding a stronger 
coupling. Thus, the more tightly coupled the travelling waves the stronger the 
magnetic field threshold that is required to transform them into travelling 
waves.

The doublet splitting at zero-field is of order $10^{-7}$, which is more than 
two orders of magnitude than the coupling parameter measured in this work. 
However, this splitting is most likely due to the non-perfect cylindrical radius. 
Once the modes transform to standing waves they sample orthogonal space along 
the azimuthal dimension and will "see" a different effective radius. Only a 
difference of 2.5 $\mu$m over the 25 mm radius is required to account for this. 
The hypothesis is supported by the fact that the tolerance of the radial 
circumference in the machining process given by the manufacture is $\pm50$ 
$\mu$m.

\section{CONCLUSION}
\label{sec:Conc}

The contribution to the magnetic susceptibility (or permeability) of the 
electron paramagnetic resonance of Fe$^{3+}$ ions in sapphire at 12.03 GHz 
(between spin states $|1/2>$ to $|3/2>$) with respect to a DC axial magnetic 
field has been measured from precision measurements of Q-factor and frequency of 
a whispering gallery mode doublet pair at 4.2 K. The form of the susceptibility 
was shown to be predominantly gyrotropic causing non-reciprocal response of the 
measured resonant modes. We showed that the gyrotropic response converted the 
standing waves to travelling waves due to the non-reciprocal response.

\section*{ACKNOWLEDGMENTS}
\label{sec:Aknowledge}

This work was supported by the Centre National d'\'Etudes Spatiales (CNES), 
the Agence Nationale pour la Recherche (ANR), the Australian Research Council 
(ARC) and travelling support from FAST program from Egide and the International 
Science Linkages program from DEST.

%

\bibliographystyle{apsrmp}


\begin{thebibliography}{42}
\expandafter\ifx\csname natexlab\endcsname\relax\def\natexlab#1{#1}\fi
\expandafter\ifx\csname bibnamefont\endcsname\relax
  \def\bibnamefont#1{#1}\fi
\expandafter\ifx\csname bibfnamefont\endcsname\relax
  \def\bibfnamefont#1{#1}\fi
\expandafter\ifx\csname citenamefont\endcsname\relax
  \def\citenamefont#1{#1}\fi
\expandafter\ifx\csname url\endcsname\relax
  \def\url#1{\texttt{#1}}\fi
\expandafter\ifx\csname urlprefix\endcsname\relax\def\urlprefix{URL }\fi
\providecommand{\bibinfo}[2]{#2}
\providecommand{\eprint}[2][]{\url{#2}}

\bibitem[{\citenamefont{Jones} \emph{et~al.}(1988)\citenamefont{Jones,
  Blair, and Buckingham}}]{JonesBlair1988el}
\bibinfo{author}{\bibnamefont{S. K.}, \bibfnamefont{Jones}},
  \bibinfo{author}{\bibfnamefont{D. G.} \bibnamefont{Blair}},
  \bibinfo{author}{\bibfnamefont{M. J.}~\bibnamefont{Buckingham}}, and
  \bibinfo{author}{\bibfnamefont{H.}~\bibnamefont{Ooguri}}, \bibinfo{year}{1988},
  \bibinfo{journal}{Electronics Letters} \textbf{\bibinfo{volume}{346}},
  \bibinfo{pages}{346}.

\bibitem[{\citenamefont{Mann} \emph{et~al.}(1992)\citenamefont{Mann,
  Giles, and Blair}}]{Mann1992jpDap}
\bibinfo{author}{\bibnamefont{A. G.}, \bibfnamefont{Mann}},
  \bibinfo{author}{\bibfnamefont{A. J.} \bibnamefont{Giles}}, and
  \bibinfo{author}{\bibfnamefont{D. G.}~\bibnamefont{Blair}}, \bibinfo{year}{1992},
  \bibinfo{journal}{Journal of Physics D-Applied Physics} \textbf{\bibinfo{volume}{25}},
  \bibinfo{pages}{1105}.


\bibitem[{\citenamefont{Luiten} \emph{et~al.}(1996)\citenamefont{Luiten,
  Mann, and Blair}}]{Luiten1996jpDap}
\bibinfo{author}{\bibnamefont{A. N.}, \bibfnamefont{Luiten}},
  \bibinfo{author}{\bibfnamefont{A. G.} \bibnamefont{Mann}}, and
  \bibinfo{author}{\bibfnamefont{D. G.}~\bibnamefont{Blair}}, \bibinfo{year}{1996},
  \bibinfo{journal}{Journal of Physics D-Applied Physics} \textbf{\bibinfo{volume}{29}},
  \bibinfo{pages}{2082}.


\bibitem[{\citenamefont{Kovacich} \emph{et~al.}(1997)\citenamefont{Kovacich,
  Mann, and Blair}}]{Kovacich1997jpDap}
\bibinfo{author}{\bibnamefont{R. P.}, \bibfnamefont{Kovacich}},
  \bibinfo{author}{\bibfnamefont{A. G.} \bibnamefont{Mann}}, and
  \bibinfo{author}{\bibfnamefont{D. G.}~\bibnamefont{Blair}}, \bibinfo{year}{1997},
  \bibinfo{journal}{Journal of Physics D-Applied Physics} \textbf{\bibinfo{volume}{30}},
  \bibinfo{pages}{3146}.

\bibitem[{\citenamefont{Giles} \emph{et~al.}(1990)\citenamefont{Giles,
  Mann, and Jones}}]{Giles1990PhysicaB}
\bibinfo{author}{\bibnamefont{A. J.}, \bibfnamefont{Giles}},
  \bibinfo{author}{\bibfnamefont{A. G.} \bibnamefont{Mann}}, and
  \bibinfo{author}{\bibfnamefont{S. K.}~\bibnamefont{Jones}}, \bibinfo{year}{1990},
  \bibinfo{journal}{Physica B} \textbf{\bibinfo{volume}{165}},
  \bibinfo{pages}{145}.


\bibitem[{\citenamefont{Bourgeois} \emph{et~al.}(2004)\citenamefont{Bourgeois,
  Lardet, Kersale, Bazin, Chaubet, and Giordano}}]{Pyb2004el}
\bibinfo{author}{\bibnamefont{P. Y.}, \bibfnamefont{Bourgeois}},
  \bibinfo{author}{\bibfnamefont{F.} \bibnamefont{Lardet-Vieudrin}}, 
  \bibinfo{author}{\bibfnamefont{Y.} \bibnamefont{Kersal\'e}},
  \bibinfo{author}{\bibfnamefont{N.} \bibnamefont{Bazin}},
  \bibinfo{author}{\bibfnamefont{M.} \bibnamefont{Chaubet}}, and
  \bibinfo{author}{\bibfnamefont{V.}~\bibnamefont{Giordano}}, \bibinfo{year}{2004},
  \bibinfo{journal}{Electronics Letters} \textbf{\bibinfo{volume}{605}},
  \bibinfo{pages}{605}.


\bibitem[{\citenamefont{Hartnett} \emph{et~al.}(2006)\citenamefont{Hartnett,
  Loke, and Ivanov}}]{Hartenett2006apl}
\bibinfo{author}{\bibnamefont{J. GY.}, \bibfnamefont{Hartnett}},
  \bibinfo{author}{\bibfnamefont{C. R.} \bibnamefont{Loke}}, and
  \bibinfo{author}{\bibfnamefont{E. N.} \bibnamefont{Ivanov}},\bibinfo{year}{2006},
  \bibinfo{journal}{Applied Physics Letters} \textbf{\bibinfo{volume}{89}},
  \bibinfo{pages}{203513}.

\bibitem[{\citenamefont{Locke} \emph{et~al.}(2008)\citenamefont{Loke,
  Ivanov, and Hartnett}}]{Locke2006rsi}
\bibinfo{author}{\bibfnamefont{C. R.} \bibnamefont{Loke}}, 
  \bibinfo{author}{\bibfnamefont{E. N.} \bibnamefont{Ivanov}}, and
  \bibinfo{author}{\bibnamefont{J. G.}, \bibfnamefont{Hartnett}}, \bibinfo{year}{2008},
  \bibinfo{journal}{Review of Scientific Instruments} \textbf{\bibinfo{volume}{79}},
  \bibinfo{pages}{051301}.


\bibitem[{\citenamefont{Hartnett} \emph{et~al.}(1999)\citenamefont{Hartnett,
  Tobar, and Mann}}]{Hartnet1999uffc}
\bibinfo{author}{\bibfnamefont{J. G.} \bibnamefont{Hartnett}}, 
  \bibinfo{author}{\bibfnamefont{M. E.} \bibnamefont{Tobar}}, and
  \bibinfo{author}{\bibnamefont{A. G.}, \bibfnamefont{Mann}}, \bibinfo{year}{1999},
  \bibinfo{journal}{IEEE Transactions on Ultrasonics Ferroelectrics and Frequency Control} \textbf{\bibinfo{volume}{46}},
  \bibinfo{pages}{993}.


\bibitem[{\citenamefont{Luiten} \emph{et~al.}(2001)\citenamefont{Luiten}}]{Hartnet2000fmc}
\bibinfo{author}{\bibfnamefont{A. N.} \bibnamefont{ Luiten (Ed.)}}, \bibinfo{year}{2001},
  \bibinfo{journal}{Frequency Measurement and Control, Topics Appl. Phys.} \textbf{\bibinfo{volume}{79}},
  \bibinfo{pages}{67-91}.


\bibitem[{\citenamefont{Hartnett} \emph{et~al.}(2007)\citenamefont{Hartnett,
  Tobar, and Le Floch}}]{Hartnet2007prB}
\bibinfo{author}{\bibfnamefont{J. G.} \bibnamefont{Hartnett}}, 
  \bibinfo{author}{\bibfnamefont{M. E.} \bibnamefont{Tobar}}, and
  \bibinfo{author}{\bibnamefont{J. M.}, \bibfnamefont{Le Floch}}, \bibinfo{year}{2007},
  \bibinfo{journal}{Physical Review B} \textbf{\bibinfo{volume}{75}},
  \bibinfo{pages}{024415}.

\bibitem[{\citenamefont{Hartnett} \emph{et~al.}(2001)\citenamefont{Hartnett,
  Tobar, and Krupka}}]{Hartnet2001jpDap}
\bibinfo{author}{\bibfnamefont{J. G.} \bibnamefont{Hartnett}}, 
  \bibinfo{author}{\bibfnamefont{M. E.} \bibnamefont{Tobar}}, and
  \bibinfo{author}{\bibnamefont{J.}, \bibfnamefont{Krupka}}, \bibinfo{year}{2001},
  \bibinfo{journal}{Journal of Physics D-Applied Physics} \textbf{\bibinfo{volume}{34}},
  \bibinfo{pages}{959}.

\bibitem[{\citenamefont{Bourgeois} \emph{et~al.}(2005)\citenamefont{Bourgeois,
  Bazin, and Kersal\'e}}]{Pyb2005apl}
\bibinfo{author}{\bibfnamefont{P. Y.} \bibnamefont{Bourgeois}}, 
  \bibinfo{author}{\bibfnamefont{N.} \bibnamefont{Bazin}}, and
  \bibinfo{author}{\bibnamefont{Y.}, \bibfnamefont{Kersal\'e}}, \bibinfo{year}{2005},
  \bibinfo{journal}{Applied Physics Letters} \textbf{\bibinfo{volume}{87}},
  \bibinfo{pages}{224104}.

\bibitem[{\citenamefont{Bourgeois} \emph{et~al.}(2006)\citenamefont{Bourgeois,
  Oxborrow, and Tobar}}]{Pyb2006ijmpB}
\bibinfo{author}{\bibfnamefont{P. Y.} \bibnamefont{Bourgeois}}, 
  \bibinfo{author}{\bibfnamefont{M.} \bibnamefont{Oxborrow}}, and
  \bibinfo{author}{\bibnamefont{M. E.}, \bibfnamefont{Tobar}}, \bibinfo{year}{2006},
  \bibinfo{journal}{International Journal of Modern Physics B} \textbf{\bibinfo{volume}{20}},
  \bibinfo{pages}{1606}.

\bibitem[{\citenamefont{Benmessai} \emph{et~al.}(2007)\citenamefont{Benmessai, Bourgeois,
  Kersal\'e, Bazin, Tobar, Hartnett, Oxborrow, Giordano}}]{Benmessai2007el}
\bibinfo{author}{\bibfnamefont{K.} \bibnamefont{Benmessai}}, 
  \bibinfo{author}{\bibfnamefont{P. Y.} \bibnamefont{Bourgeois}}, 
  \bibinfo{author}{\bibfnamefont{Y.} \bibnamefont{Kersal\'e}}, 
  \bibinfo{author}{\bibfnamefont{N.} \bibnamefont{Bazin}}, 
  \bibinfo{author}{\bibfnamefont{M. E.} \bibnamefont{Tobar}}, 
  \bibinfo{author}{\bibfnamefont{J. G.} \bibnamefont{Hartnett}}, 
  \bibinfo{author}{\bibfnamefont{M.} \bibnamefont{Oxborrow}}, and
  \bibinfo{author}{\bibnamefont{V.}, \bibfnamefont{Giordano}}, \bibinfo{year}{2007},
  \bibinfo{journal}{Electronics Letters} \textbf{\bibinfo{volume}{43}},
  \bibinfo{pages}{1436}.

\bibitem[{\citenamefont{Benmessai} \emph{et~al.}(2007)\citenamefont{Benmessai, Creedon,
  and Tobar}}]{Benmessai2008prl}
\bibinfo{author}{\bibfnamefont{K.} \bibnamefont{Benmessai}}, 
  \bibinfo{author}{\bibfnamefont{D. L.} \bibnamefont{Creedon}}, and
  \bibinfo{author}{\bibfnamefont{M. E.} \bibnamefont{Tobar}}, \bibinfo{year}{2008},
  \bibinfo{journal}{Physical Review Letters} \textbf{\bibinfo{volume}{100}},
  \bibinfo{pages}{233901}.



\bibitem[{\citenamefont{Tobar} \emph{et~al.}(1991)\citenamefont{Tobar, and Mann}}]{TobarMann1991mtt}
\bibinfo{author}{\bibfnamefont{M. E.} \bibnamefont{Tobar}}, and
  \bibinfo{author}{\bibfnamefont{A. G.} \bibnamefont{Mann}}, \bibinfo{year}{1991},
  \bibinfo{journal}{IEEE Transactions on Microwave Theory and Techniques} \textbf{\bibinfo{volume}{39}},
  \bibinfo{pages}{2077}.




\bibitem[{\citenamefont{Wolf} \emph{et~al.}(2004)\citenamefont{Wolf, Tobar, and Bize}}]{Wolf2004grg}
\bibinfo{author}{\bibfnamefont{P.} \bibnamefont{Wolf}}, 
  \bibinfo{author}{\bibfnamefont{M. E.} \bibnamefont{Tobar}}, and 
  \bibinfo{author}{\bibfnamefont{S.} \bibnamefont{Bize}}, \bibinfo{year}{2004},
  \bibinfo{journal}{IEEE Transactions on Microwave Theory and Techniques} \textbf{\bibinfo{volume}{36}},
  \bibinfo{pages}{2004}.


\bibitem[{\citenamefont{Weiss} \emph{et~al.}(1995)\citenamefont{Weiss, Sandoghdar, and Hare}}]{Weiss1995ol}
\bibinfo{author}{\bibfnamefont{D. S.} \bibnamefont{Weiss}}, 
  \bibinfo{author}{\bibfnamefont{V.} \bibnamefont{Sandoghdar}}, and 
  \bibinfo{author}{\bibfnamefont{J.} \bibnamefont{Hare}}, \bibinfo{year}{1995},
  \bibinfo{journal}{Optics Letters} \textbf{\bibinfo{volume}{20}},
  \bibinfo{pages}{1835}.

\bibitem[{\citenamefont{Kippenberg} \emph{et~al.}(2002)\citenamefont{Kippenberg, Spillane, and Vahala}}]{Kippenberg2002ol}
\bibinfo{author}{\bibfnamefont{T. J.} \bibnamefont{Kippenberg}}, 
  \bibinfo{author}{\bibfnamefont{S. M.} \bibnamefont{Spillane}}, and 
  \bibinfo{author}{\bibfnamefont{K. J.} \bibnamefont{Vahala}}, \bibinfo{year}{2002},
  \bibinfo{journal}{Optics Letters} \textbf{\bibinfo{volume}{27}},
  \bibinfo{pages}{1669}.


\bibitem[{\citenamefont{Gorodetsky} \emph{et~al.}(2000)\citenamefont{Gorodetsky, Pryamikov, and Ilchenko}}]{Gorodetsky2000josaBop}
\bibinfo{author}{\bibfnamefont{M. L.} \bibnamefont{Gorodetsky}}, 
  \bibinfo{author}{\bibfnamefont{A. D.} \bibnamefont{Pryamikov}}, and 
  \bibinfo{author}{\bibfnamefont{V. S.} \bibnamefont{VIlchenko}}, \bibinfo{year}{2000},
  \bibinfo{journal}{Journal of the Optical Society of America B-Optical Physics} \textbf{\bibinfo{volume}{17}},
  \bibinfo{pages}{1051}.

\bibitem[{\citenamefont{Krupka} \emph{et~al.}(1996)\citenamefont{Krupka, Blondy, and Cros}}]{Krupka1996mtt}
\bibinfo{author}{\bibfnamefont{J.} \bibnamefont{Krupka}}, 
  \bibinfo{author}{\bibfnamefont{P.} \bibnamefont{Blondy}}, and 
  \bibinfo{author}{\bibfnamefont{D.} \bibnamefont{Cross}}, \bibinfo{year}{1996},
  \bibinfo{journal}{IEEE Transactions on Microwave Theory and Techniques} \textbf{\bibinfo{volume}{44}},
  \bibinfo{pages}{1097}.


\bibitem[{\citenamefont{Krupka} \emph{et~al.}(1996)\citenamefont{Krupka, and Geyer}}]{Krupka1996mttbis}
\bibinfo{author}{\bibfnamefont{J.} \bibnamefont{Krupka}}, and
  \bibinfo{author}{\bibfnamefont{R. G.} \bibnamefont{Geyer}}, \bibinfo{year}{1996},
  \bibinfo{journal}{IEEE Transactions on Microwave Theory and Techniques} \textbf{\bibinfo{volume}{32}},
  \bibinfo{pages}{1924}.


\bibitem[{\citenamefont{Tobar} \emph{et~al.}(1997)\citenamefont{Tobar, Krupka, and Ivanov}}]{Tobar1997jpDap}
\bibinfo{author}{\bibfnamefont{M. E.} \bibnamefont{Tobar}}, 
  \bibinfo{author}{\bibfnamefont{J.} \bibnamefont{Krupka}}, and
  \bibinfo{author}{\bibfnamefont{E. N.} \bibnamefont{Ivanov}}, \bibinfo{year}{1997},
  \bibinfo{journal}{Journal of Physics D-Applied Physics} \textbf{\bibinfo{volume}{30}},
  \bibinfo{pages}{2770}.


\bibitem[{\citenamefont{Krupka} \emph{et~al.}(1996)\citenamefont{Krupka, Derzakowski, and Tobar}}]{Krupka1999mst}
\bibinfo{author}{\bibfnamefont{J.} \bibnamefont{Krupka}}, 
  \bibinfo{author}{\bibfnamefont{K.} \bibnamefont{Derzakowski}}, and
  \bibinfo{author}{\bibfnamefont{M. E.} \bibnamefont{Tobar}}, \bibinfo{year}{1999},
  \bibinfo{journal}{Measurement Science \& Technology} \textbf{\bibinfo{volume}{10}},
  \bibinfo{pages}{387}.


\bibitem[{\citenamefont{Krupka} \emph{et~al.}(1996)\citenamefont{Krupka, Derzakowski, and Abramowicz}}]{Krupka1999mtt}
\bibinfo{author}{\bibfnamefont{J.} \bibnamefont{Krupka}}, 
  \bibinfo{author}{\bibfnamefont{K.} \bibnamefont{Derzakowski}}, and
  \bibinfo{author}{\bibfnamefont{A.} \bibnamefont{Abramowicz}}, \bibinfo{year}{1999},
  \bibinfo{journal}{IEEE Transactions on Microwave Theory and Techniques} \textbf{\bibinfo{volume}{47}},
  \bibinfo{pages}{752}.

\bibitem[{\citenamefont{Gurevich} \emph{et~al.}(1960)\citenamefont{Gurevich}}]{Gurevich1960sppl}
\bibinfo{author}{\bibfnamefont{A. G.} \bibnamefont{Gurevich1960sppl}}, \bibinfo{year}{1960},
  \bibinfo{title}{Ferrites at microwave frequencies}, 
  \bibinfo{journal}{State Press for Physicomathematical Literature, Moscow}.

\bibitem[{\citenamefont{Tobar} \emph{et~al.}(2003)\citenamefont{Tobar, and Hartnett}}]{Tobar2003prD}
\bibinfo{author}{\bibfnamefont{M. E.} \bibnamefont{Tobar}}, and
  \bibinfo{author}{\bibfnamefont{J. G.} \bibnamefont{Hartnett}}, \bibinfo{year}{2003},
  \bibinfo{journal}{Physical Review D} \textbf{\bibinfo{volume}{67}},
  \bibinfo{pages}{062001}.


\bibitem[{\citenamefont{Benabid} \emph{et~al.}(2000)\citenamefont{Benabid, Notcutt, and Loriette}}]{Benabid2000}
\bibinfo{author}{\bibfnamefont{F.} \bibnamefont{Benabid}}, 
  \bibinfo{author}{\bibfnamefont{M.} \bibnamefont{Notcutt}}, and
  \bibinfo{author}{\bibfnamefont{V.} \bibnamefont{Loriette}}, \bibinfo{year}{2000},
  \bibinfo{journal}{Journal of Physics D-Applied Physics} \textbf{\bibinfo{volume}{33}},
  \bibinfo{pages}{589}.


\bibitem[{\citenamefont{Bourgeois} \emph{et~al.}(2005)\citenamefont{Bourgeois and Giordano}}]{pyb2005mtt}
\bibinfo{author}{\bibfnamefont{P. Y.} \bibnamefont{Bourgeois}}, and
  \bibinfo{author}{\bibfnamefont{V.} \bibnamefont{Giordano}}, \bibinfo{year}{2005},
  \bibinfo{journal}{IEEE Transactions on Microwave Theory and Techniques} \textbf{\bibinfo{volume}{53}},
  \bibinfo{pages}{3185}.

\bibitem[{\citenamefont{Mazzei} \emph{et~al.}(2007)\citenamefont{Mazzei, Gotezinger, and Menezes}}]{Mazzei2007prl}
\bibinfo{author}{\bibfnamefont{A.} \bibnamefont{Mazzei}}, 
  \bibinfo{author}{\bibfnamefont{S.} \bibnamefont{Gotezinger}}, and 
  \bibinfo{author}{\bibfnamefont{L de S.} \bibnamefont{Menezes}}, \bibinfo{year}{2007},
  \bibinfo{journal}{Physical Review Letters} \textbf{\bibinfo{volume}{99}},
  \bibinfo{pages}{173603}.


\bibitem[{\citenamefont{Tobar} \emph{et~al.}(1991)\citenamefont{Tobar, and Blair}}]{Tobar1991mtt}
\bibinfo{author}{\bibfnamefont{M. E.} \bibnamefont{Tobar}}, and 
  \bibinfo{author}{\bibfnamefont{D. G.} \bibnamefont{Blair}}, \bibinfo{year}{1991},
  \bibinfo{journal}{IEEE Transactions on Ultrasonics Ferroelectrics and Frequency Control} \textbf{\bibinfo{volume}{39}},
  \bibinfo{pages}{1582}.



\bibitem[{\citenamefont{Tobar} \emph{et~al.}(1993)\citenamefont{Tobar}}]{Tobar1993jpDap}
\bibinfo{author}{\bibfnamefont{M. E.} \bibnamefont{Tobar}}, \bibinfo{year}{1991},
  \bibinfo{journal}{Journal of Physics D-Applied Physics} \textbf{\bibinfo{volume}{26}},
  \bibinfo{pages}{2022}.


\bibitem[{\citenamefont{Kornienko} \emph{et~al.}(1961)\citenamefont{Kornienko, and Prokhorov}}]{Kornienko1961}
\bibinfo{author}{\bibfnamefont{L. S.} \bibnamefont{Kornienko}}, and 
  \bibinfo{author}{\bibfnamefont{A. M.} \bibnamefont{Prokhorov}}, \bibinfo{year}{1961},
  \bibinfo{journal}{Sov. Phys.-JETP} \textbf{\bibinfo{volume}{13}},
  \bibinfo{pages}{1120}.






\end{thebibliography}

\end{document}